\documentclass[aps,pra,floatfix,twocolumn,amsmath,amssymb,showpacs]{revtex4-1}

\usepackage{graphicx}
\usepackage{pst-all,pstricks-add}

\newcommand{\ket}[1]{\lvert #1 \rangle}           %Create a ket with an argument
\newcommand{\bra}[1]{\langle #1 \lvert}           %Create a Bra with an argument

\newcommand{\tr}{\mbox{\scriptsize\sf T}}
\newcommand{\cS}{\mathcal{S}}
\newcommand{\bI}{\mathbf{I}}
\newcommand{\bl}{\mathbf{1}}
\newcommand{\bJ}{\mathbf{J}}
\newcommand{\bL}{\mathbf{L}}
\newcommand{\bD}{\mathbf{D}}
\newcommand{\bOmega}{\mathbf{\Omega}}
\newcommand{\bA}{\mathbf{A}}
\newcommand{\ba}{\mathbf{a}}
\newcommand{\bb}{\mathbf{b}}
\newcommand{\br}{\mathbf{r}}
\newcommand{\bz}{\mathbf{0}}
\newcommand{\bx}{\mathbf{x}}
\newcommand{\by}{\mathbf{y}}
\newcommand{\bZ}{\mathbf{Z}}
\newcommand{\bR}{\mathbf{R}}
\newcommand{\bM}{\mathbf{M}}
\newcommand{\bS}{\mathbf{S}}
\newcommand{\bC}{\mathbf{C}}
\newcommand{\bT}{\mathbf{T}}
\newcommand{\bU}{\mathbf{U}}
\newcommand{\bO}{\mathbf{\Omega}}

\newcommand{\bpsi}{\mathbf{\psi}}
\newcommand{\bLambda}{\mathbf{\Lambda}}

\DeclareMathOperator{\Trace}{Tr}

\DeclareMathOperator{\mse}{mse}
\DeclareMathOperator{\expE}{E}

\begin{document}

\author{Michael Frey}
\affiliation{Department of Mathematics, Bucknell University, Lewisburg, PA 17837, USA}
\email{mfrey@bucknell.edu}
\thanks{Author to whom correspondence should be addressed.}

\author{David Collins}
\affiliation{Department of Physical and Environmental Sciences, Mesa State College, 1100 North Avenue, Grand Junction, CO 81501, USA}
\email{dacollin@mesastate.edu}

\author{Karl Gerlach}
\affiliation{ITT Corp., 2560 Huntington Avenue, Alexandria, VA 22303, USA}

\title{Probing the qudit depolarizing channel}

\begin{abstract}
For the quantum depolarizing channel with any finite dimension, we compare three schemes for channel identification: unentangled probes, probes maximally entangled with an external ancilla, and maximally entangled probe pairs. This comparison includes cases where the ancilla is itself depolarizing and where the probe is circulated back through the channel before measurement. Compared on the basis of (quantum Fisher) information gained per channel use, we find broadly that entanglement with an ancilla dominates the other two schemes, but only if entanglement is cheap relative to the cost per channel use and only if the external ancilla is well shielded from depolarization. We arrive at these results by a relatively simple analytical means. A separate, more complicated analysis for partially entangled probes shows for the qudit depolarizing channel that any amount of probe entanglement is advantageous and that the greatest advantage comes with maximal entanglement.
\end{abstract}


\maketitle

%%%%%%%%%%%%%%%%%%%%%%%%%%%%%%%%%%%%%%%%%%%%%%%%%%%%%%%%%%%%%%%%%%%%%%%%%%%%%%%%%%%%%%%%%%%%%%%%%%%%%%%%%
%%%%%%%%%%%%%%%%%%%                                                                       %%%%%%%%%%%%%%%
%%%%%%%%%%%%%%%%%%%         Begin section                                                 %%%%%%%%%%%%%%%
%%%%%%%%%%%%%%%%%%%                                                                       %%%%%%%%%%%%%%%
%%%%%%%%%%%%%%%%%%%%%%%%%%%%%%%%%%%%%%%%%%%%%%%%%%%%%%%%%%%%%%%%%%%%%%%%%%%%%%%%%%%%%%%%%%%%%%%%%%%%%%%%%

\section{Introduction}
\label{sec:intro}

A quantum operation, or channel, $\Gamma$ is a trace-preserving, completely positive
map for representing the change of state of a quantum system. Quantum channels are the
fundamental building blocks of quantum information processing. In any physical
implementation, though, the channel is usually not fully known and must be determined
experimentally by observing its effect on prepared quantum systems, or probes. This is
the problem of quantum channel identification
\cite{fujiwara01,sasaki02,fujiwara03,fujiwara04,ballester04,frey10a,frey10b}, also known
as quantum process tomography \cite{nielsen00}, and a major challenge is to determine
the optimal quantum state and measurement for the channel probe. In particular, the
probe may be entangled with an ancilla system, or with other probes.

\begin{figure}[h]
\includegraphics[scale=.68]{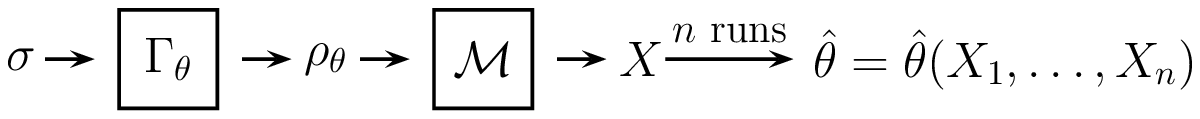}
\caption{Set-up for probing a depolarizing channel $\Gamma_{\theta}$ by a probe system prepared in state $\sigma$.
\label{fig:channel}}
\end{figure}

Quantum channel identification is statistical by nature and is generally formulated
as a parameter estimation problem: the unknown channel $\Gamma_\theta$ is supposed to
belong to a parametric family $\{\Gamma_\theta, \, \theta \in \Theta\}$ of channels,
and we identify the channel by estimating $\theta$. The general scheme for doing this
is to prepare the channel input probe in a chosen quantum state $\sigma$, make a quantum
measurement ${\cal M}$ of the channel output $\rho_\theta=\Gamma_\theta[\sigma]$,
and base the estimation of $\theta$ on the measurement's registered result $X$. This
process (see Fig.~\ref{fig:channel}) is repeated to obtain $n$ independent, identically
distributed measurement outcomes $X_1, \ldots, X_n$, and then $\theta$ is estimated
by an estimator $\hat\theta=\hat\theta(X_1, \ldots, X_n)$. This arrangement, in which
a prepared quantum system is used to probe a quantum channel, is standard for channel
identification, it is common in interferometry \cite{shapiro91}, and it occurs generally
whenever a quantum measurement is made indirectly by observing a change in an ancillary
part of the measurement apparatus.

We study the problem of identifying the qudit (dimension $d < \infty$) depolarizing channel. This channel is a standard
model of quantum noise, it is analytically tractable, and it is the basis of investigation in a range of contexts \cite{frey10b,king03,ballester05,slimen07,boixo08,daems07}. Our study of the qudit depolarizing channel proceeds most directly from the works of Fujiwara and Sasaki {\it et al}. Fujiwara \cite{fujiwara01} considers the qubit ($d=2$) depolarizing channel with probes prepared by three different schemes. Using a likelihood approach based around quantum Fisher information, Fujiwara obtains
comprehensive ``local/asymptotic'' optimality results for his three identification schemes, even including probes prepared in partially entangled states. Sasaki {\it et al.} \cite{sasaki02} study the more general qudit ($d<\infty$) depolarizing channel from a Bayesian perspective, comparing the same channel identification schemes as Fujiwara. Their Bayesian approach yields ``global'' optimality results for any $d<\infty$, though their results for partial entanglement are limited to $d=2$. We take the same
approach as Fujiwara but treat the general qudit depolarizing channel, introducing a common simple expression for the quantum Fisher information to address all three identification schemes treated in \cite{fujiwara01,sasaki02}. We extend this method to entanglement with an ancilla that is itself depolarizing and to a scheme we call probe re-circulation, whereby the probe is passed multiple times through the channel before measurement. By a separate analysis we show for the depolarizing channel that any
degree of entanglement is beneficial for qudits of any dimension and that maximal entanglement is most beneficial. This broad usefulness of entanglement in any dimension is known in only a few instances \cite{piani09}.

The remainder of the paper is arranged as follows. We set the stage for our work in Sec.~\ref{sec:prelim} by recalling the definition of quantum Fisher information and its role in the quantum Cram\'{e}r-Rao inequality. We then derive a general expression for the quantum Fisher information in the output of the qudit depolarizing channel. In Sec.~\ref{sec:withentangle} we apply this expression to channel identification schemes in which the qudit probe is unentangled, maximally entangled with an ancilla
qudit, or maximally entangled with another probe (the three schemes studied in \cite{fujiwara01,sasaki02}). For each of these schemes, we treat the possibility that the probe is circulated back through the channel before measurement. We further apply our expression for quantum Fisher information to a probe that is maximally entangled with an ancilla qudit that is itself depolarizing. Partially entangled probes must be addressed with a separate method; we do this in Sec.~\ref{sec:partialentangle}. We
take up the question of optimal probe measurement in Sec.~\ref{sec:measurements} and summarize our results in Sec.~\ref{sec:summary}.
 
%%%%%%%%%%%%%%%%%%%%%%%%%%%%%%%%%%%%%%%%%%%%%%%%%%%%%%%%%%%%%%%%%%%%%%%%%%%%%%%%%%%%%%%%%%%%%%%%%%%%%%%%%
%%%%%%%%%%%%%%%%%%%                                                                       %%%%%%%%%%%%%%%
%%%%%%%%%%%%%%%%%%%         End section                                                 %%%%%%%%%%%%%%%
%%%%%%%%%%%%%%%%%%%                                                                       %%%%%%%%%%%%%%%
%%%%%%%%%%%%%%%%%%%%%%%%%%%%%%%%%%%%%%%%%%%%%%%%%%%%%%%%%%%%%%%%%%%%%%%%%%%%%%%%%%%%%%%%%%%%%%%%%%%%%%%%%

%%%%%%%%%%%%%%%%%%%%%%%%%%%%%%%%%%%%%%%%%%%%%%%%%%%%%%%%%%%%%%%%%%%%%%%%%%%%%%%%%%%%%%%%%%%%%%%%%%%%%%%%%
%%%%%%%%%%%%%%%%%%%                                                                       %%%%%%%%%%%%%%%
%%%%%%%%%%%%%%%%%%%         Begin section                                                 %%%%%%%%%%%%%%%
%%%%%%%%%%%%%%%%%%%                                                                       %%%%%%%%%%%%%%%
%%%%%%%%%%%%%%%%%%%%%%%%%%%%%%%%%%%%%%%%%%%%%%%%%%%%%%%%%%%%%%%%%%%%%%%%%%%%%%%%%%%%%%%%%%%%%%%%%%%%%%%%%

\section{Preliminaries}
\label{sec:prelim}

The quantum Fisher information $J(\theta)=J[\rho_\theta]$ in the parametric output $\rho_\theta=\Gamma_\theta(\sigma)$ of a quantum channel bounds the ultimate precision of the estimation of the parameter $\theta$ attainable by quantum measurement $\cal M$ of $\rho_\theta$. The quantum information (Cram\'{e}r-Rao) inequality states for $n$ independent, identically distributed registrations of any quantum measurement $\cal M$ and any unbiased estimator $\hat\theta$ that
\begin{equation}
\mse{[\hat\theta]} \geqslant \frac{1}{nJ(\theta)}
\label{eq:crb}
\end{equation}
where $\mse{[\hat\theta]}=\expE{[(\hat\theta-\theta)^2]}$ is the mean squared error of
$\hat\theta$. The quantum Fisher information $J(\theta)$ in Eq.~\eqref{eq:crb} is
\begin{equation}
    J(\theta)=J[\rho_{\theta}]=\mbox{tr}[\rho_{\theta}L_{\theta}^2]
\end{equation}
where $L_{\theta}$ is the quantum score operator (symmetric logarithmic derivative) of $\rho_{\theta}$ defined by
\begin{equation}
\frac{L_\theta\rho_\theta+\rho_\theta L_\theta}{2}
=\partial_\theta\rho_\theta
\label{eq:score}
\end{equation}
where $\partial_\theta$ signifies differentiation. Quantum Fisher information originated with Helstrom \cite{helstrom67}; its relationship to classical Fisher information is brought out in \cite{braunstein94}; and recent presentations are \cite{paris09,chen07}. The Cram\'{e}r-Rao bound, Eq.~\eqref{eq:crb}, is asymptotically achievable so the larger $J(\theta)$ is in Eq.~\eqref{eq:crb}, the more precisely $\theta$ can potentially be estimated. We therefore interpret $J(\theta)$ to be the overall
information about $\theta$ available in a parametrically defined quantum state $\rho_{\theta}$. In this sense $J(\theta)$ serves as a quantitative measure of the relative merit of different channel identification schemes. The information inequality, Eq.~\eqref{eq:crb}, and this interpretation of quantum Fisher information, readily extends~\cite{paris09} to biased estimators, dependent and non-identically distributed registrations, vectors of channel parameters~\cite{frey09}, and transformations of
parameter.

The qudit ($d$-dimensional) depolarizing channel defined for any qudit input state $\sigma$ is
\begin{equation}
\Gamma_{\theta}(\sigma)=\frac{1-\theta}{d}\bI_d+\theta\sigma
\end{equation}
where $1-\theta$ is the probability of depolarization and $\frac{1}{d}\bI_d$ is the completely mixed qudit state. The channel
$\Gamma_{\theta}$ is completely positive for $0\leqslant \theta \leqslant 1$. In fact \cite{fujiwara99}, $\Gamma_{\theta}$ is completely positive for $-1/(d^2-1)\leqslant  \theta \leqslant 1$; we restrict our attention to $0\leqslant \theta \leqslant 1$ where $\theta$ reflects solely depolarization. We take advantage of a general transformation of parameter to simply and flexibly express the quantum Fisher information $J(\theta)$ for the qudit depolarizing channel. This formulation of $J(\theta)$ is used
in the following section to develop and compare schemes for estimating $\theta$ and identifying $\Gamma_{\theta}$. In particular, we generalize results known \cite{fujiwara01} for the qubit depolarizing channel to the setting of qudits.

We will need to consider a variety of depolarizing-type channels, whose general form is
\begin{equation}
\sigma \rightarrow \frac{1 - h(\theta)}{d} I_d + h(\theta) \sigma
\end{equation}
where $d$ is the dimension of the system and $ 0 \leqslant h(\theta) \leqslant 1$ is monotonic and has the interpretation that $1 - h(\theta)$ is a generalized depolarizing probability. Given that the probe state $\sigma$ at the input of the qudit depolarizing channel has spectral decomposition $\sigma = \sum_{i=1}^d s_i\ket{s_i}\bra{s_i}$ it follows that $\rho_{\theta} = \Gamma_{\theta}(\sigma) = \sum_{i=1}^d \lambda_i(\theta) \ket{s_i}\bra{s_i}$ where
\begin{equation}
\lambda_i(\theta) = \frac{1-h(\theta)}{d}+s_ih(\theta).
\label{eq:ev}
\end{equation}
Only the eigenvalues $\lambda_i(\theta)$ of $\rho_\theta$ depend on $\theta$; its eigenvectors $\ket{s_i}$ do not, so
$\rho_{\theta}$ is quasi-classical \cite{paris09,frey09}. In this situation $\rho_\theta$ and the score operator $L_\theta$
commute, $L_\theta$ is readily calculated, and we have
\begin{equation}
J(\theta) = \Trace{[\rho_{\theta}L_\theta^2]}
          = \Trace{[\rho_{\theta}^{-1}(\partial_\theta\rho_\theta)^2]}
          = \sum_{i=1}^d
            \frac{\lambda_i^\prime(\theta)^2}{\lambda_i(\theta)}.
\label{eq:qfi}
\end{equation}

Fujiwara \cite{fujiwara01} observed that the quantum Fisher information $J(\theta)= J[\Gamma_{\theta}(\sigma)]$ in a channel output $\Gamma_{\mu}(\sigma)$ about a channel parameter $\theta$ is maximized by a pure state $\sigma$. This observation applies in particular to the qudit depolarizing channel, so we restrict our attention here to pure probe states. Let $\left\{ \ket{s_i},\,i=1,\dots,d \right\}$ be an orthonormal basis relative to which $\sigma=|s_1\rangle\langle s_1|$. The channel output is then
\begin{equation}
\rho_\theta =    \frac{1+(d-1)h(\theta)}{d} \ket{s_1}\bra{s_1}
              +  \frac{1-h(\theta)}{d} \sum_{i=2}^d \ket{s_i}\bra{s_i}.
\label{eq:re}
\end{equation}
Reading the eigenvalues of $\rho_\theta$ from Eq.~\eqref{eq:re}, we find that the quantum Fisher information Eq.~\eqref{eq:qfi} in $\rho_\theta$ is
\begin{equation}
J(\theta)= \frac{(h^\prime(\theta))^2}{(1-h(\theta))
\left( h(\theta) + \frac{1}{d-1}\right)}.
\label{eq:qfi2}
\end{equation}
This expression for $J(\theta)$ applies to all the qudit depolarizing channel identification schemes of the next section.

%%%%%%%%%%%%%%%%%%%%%%%%%%%%%%%%%%%%%%%%%%%%%%%%%%%%%%%%%%%%%%%%%%%%%%%%%%%%%%%%%%%%%%%%%%%%%%%%%%%%%%%%%
%%%%%%%%%%%%%%%%%%%                                                                       %%%%%%%%%%%%%%%
%%%%%%%%%%%%%%%%%%%         End section                                                 %%%%%%%%%%%%%%%
%%%%%%%%%%%%%%%%%%%                                                                       %%%%%%%%%%%%%%%
%%%%%%%%%%%%%%%%%%%%%%%%%%%%%%%%%%%%%%%%%%%%%%%%%%%%%%%%%%%%%%%%%%%%%%%%%%%%%%%%%%%%%%%%%%%%%%%%%%%%%%%%%

%%%%%%%%%%%%%%%%%%%%%%%%%%%%%%%%%%%%%%%%%%%%%%%%%%%%%%%%%%%%%%%%%%%%%%%%%%%%%%%%%%%%%%%%%%%%%%%%%%%%%%%%%
%%%%%%%%%%%%%%%%%%%                                                                       %%%%%%%%%%%%%%%
%%%%%%%%%%%%%%%%%%%         Begin section                                                 %%%%%%%%%%%%%%%
%%%%%%%%%%%%%%%%%%%                                                                       %%%%%%%%%%%%%%%
%%%%%%%%%%%%%%%%%%%%%%%%%%%%%%%%%%%%%%%%%%%%%%%%%%%%%%%%%%%%%%%%%%%%%%%%%%%%%%%%%%%%%%%%%%%%%%%%%%%%%%%%%

\section{Probing with entanglement}
\label{sec:withentangle}

We first consider channel identification in which the qudit probe is prepared in an unentangled pure state
$\sigma$---we call this scheme O. The probe is applied to the depolarizing channel with depolarizing probability $1-\theta$;
i.e., $h(\theta)=\theta$. We have immediately from Eq.~\eqref{eq:qfi2} that the quantum Fisher information available in the channel
output about $\theta$ is
\begin{equation}
J_{\mathrm{O}}(\theta)
   = \frac{1}{(1-\theta)\left(\theta+\frac{1}{d-1}\right)}
\label{eq:sep}
\end{equation}
with no input entanglement. This quantum Fisher information is the same for any pure input state $\sigma$, because the qudit depolarizing channel is unitarily invariant and $J[\rho_\theta]=J[\bU\rho_\theta\bU^{\dagger}]$ for any unitary transformation $\bU$.

We now consider an externally entangled probe. Specifically, in scheme E one of a maximally entangled pair of qudits is passed
through $\Gamma_\theta$ with depolarizing probability $1-\theta$; again $h(\theta)=\theta$. The state of the maximally entangled
qudit pair is $\sigma=\ket{\mu}\bra{\mu}$ with
\begin{equation}
 \ket{\mu} = \frac{\ket{s_1t_1}+\dots+\ket{s_dt_d}}{\sqrt{d}}
\label{eq:mu}
\end{equation}
where $\left\{ \ket{s_i} \right\}$ and $\left\{ \ket{t_i} \right\}$ are orthonormal bases for each qubit. The channel output
$\rho_\theta=(\bI\otimes\Gamma_\theta)(\sigma)$ is, after some calculation,
\begin{equation}
\rho_\theta = \frac{1-\theta}{d^2}\bI_{d^2} +\theta\sigma \; .
\label{eq:ente}
\end{equation}
This output is identically that of a $d^2$-dimensional qudit depolarizing channel with depolarizing probability $1-\theta$
and a probe prepared in a pure state. From Eq.~\eqref{eq:sep}, then, the quantum Fisher information in Eq.~\eqref{eq:ente} about $\theta$
obtained by a maximally entangled probe is
\begin{equation}
J_{\mathrm{E}}(\theta)
= \frac{1}{(1-\theta)\left(\theta+\frac{1}{d^2-1}\right)},
\label{eq:infe}
\end{equation}
and the information gained using a maximally entangled probe is
\begin{equation}
J_{\mathrm{E}}(\theta) - J_{\mathrm{O}}(\theta)
= \frac{d(d-1)(1-\theta)}{(1-\theta+d\theta)(1-\theta+d^2\theta)}.
\end{equation}
This gain is positive for all $d<\infty$ and all $\theta < 1$.

One might entangle two qudits and use {\it both} as probes as a scheme for channel identification. In scheme B two qudits in
the maximally entangled state $\sigma=\ket{\mu}\bra{\mu}$, with $\ket{\mu}$ as in Eq.~\eqref{eq:mu}, are each passed through the qudit depolarizing channel. This yields the channel output $\rho_\theta= (\Gamma_\theta\otimes\Gamma_\theta)(\sigma)$, in which case
\begin{equation}
\rho_\theta = \frac{1-\theta^2}{d^2}\bI_{d^2} +\theta^2\sigma.
\label{eq:entb}
\end{equation}
This is identically the parametric family of output states that would result from passing a $d^2$-dimensional unentangled qudit in the pure state $\sigma$ through a channel with depolarizing probability $1-h(\theta)$ where $h(\theta)=\theta^2$. So from Eq.~\eqref{eq:qfi2}, the information {\it per channel use} in Eq.~\eqref{eq:entb} about $\theta$ is
\begin{equation}
J_{\mathrm{B}}(\theta)
= \frac{2\theta^2}{(1-\theta^2)\left(\theta^2+\frac{1}{d^2-1}\right)}.
\label{eq:infb}
\end{equation}

Our three channel identification schemes O, E, and B are just those considered in \cite{fujiwara01,sasaki02}. These schemes can be
compared in different ways based on the quantum Fisher informations $J_{\mathrm{O}}(\theta)$, $J_{\mathrm{E}}(\theta)$ and $J_{\mathrm{B}}(\theta)$. The three schemes use different resources to different degrees, suggesting different accountings. Schemes O and E involve just one channel use; scheme B requires two uses. Schemes E and B use entanglement, whereas scheme O does not. Schemes E and B involve a dimensional extension of the channel (with a more complex measurement); scheme O does not. Both
Fujiwara \cite{fujiwara01} and Sasaki {\it et al.} \cite{sasaki02}, compared these schemes on the basis of channel extension. Our $J_{\mathrm{O}}(\theta)$, $J_{\mathrm{E}}(\theta)$, and $J_{\mathrm{B}}(\theta)$ are information gained per channel use. Thus Fujiwara effectively compared $2J_{\mathrm{O}}(\theta)$, $J_{\mathrm{E}}(\theta)$ and $2J_{\mathrm{B}}(\theta)$. He found by his accounting that no one scheme uniformly dominated the others; each scheme was optimal for different
$\theta$.

\begin{figure}[h]
  \includegraphics[scale=0.75]{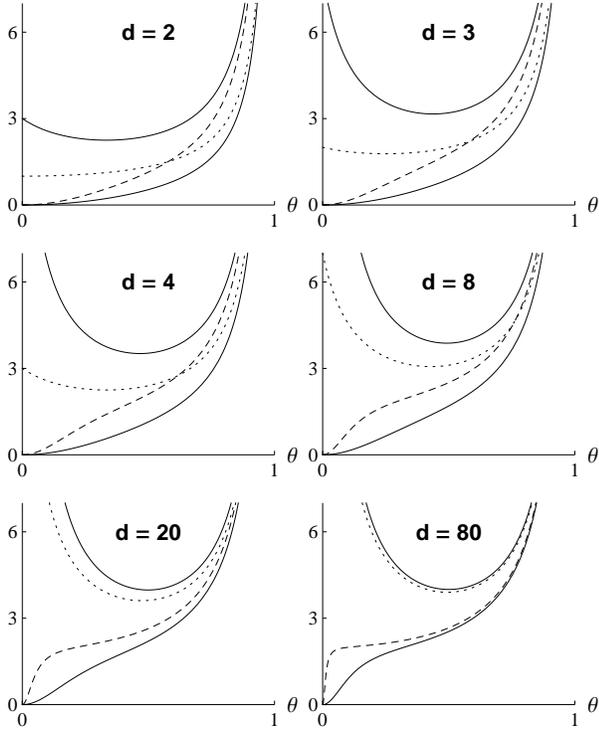}
  \caption{Quantum Fisher information per channel use: Scheme E (upper solid line), Scheme O (dotted), Scheme B (dashed), Scheme O,2 (solid lower).
  \label{fig:fig2}}
 \end{figure}

We compare our identification schemes O, E, and B for the general qudit depolarizing channel on the basis of quantum Fisher information per channel use. This comparison emphasizes the essential cost of the channel to estimation. From a resource perspective this seems most natural, because if channel use is not relatively costly, we would always just probe the channel by scheme O with as many probes as needed to achieve the desired precision. Comparing quantum Fisher information per channel use, we find
that
\begin{subequations}
\begin{eqnarray}
J_{\mathrm{E}}(\theta) & \geqslant & J_{\mathrm{O}}(\theta) \quad \textrm{and}  \\
J_{\mathrm{E}}(\theta) & \geqslant & J_{\mathrm{B}}(\theta)
\end{eqnarray}
\end{subequations}
for all $\theta$ and all channel dimensions $d$, as shown in Fig.~\ref{fig:fig2}. 

\begin{figure}[h]
  \includegraphics[scale=0.90]{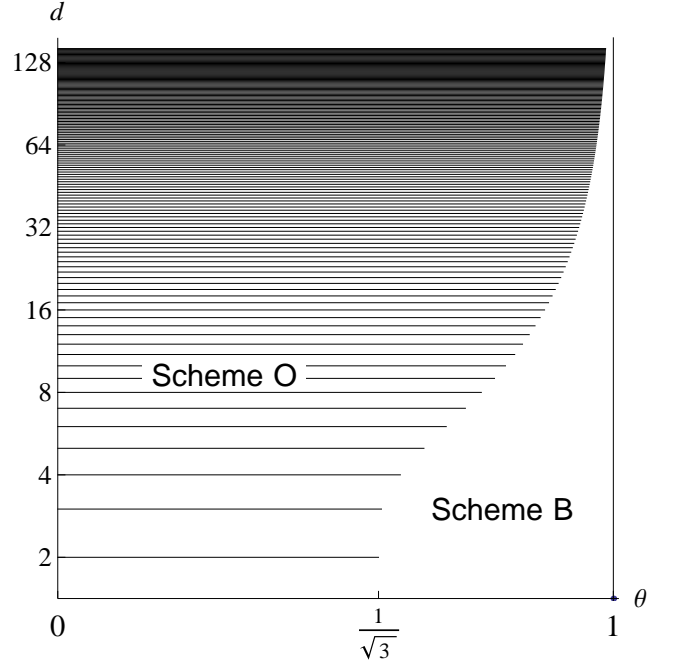}
  \caption{Unentangled probes (scheme O) versus maximally entangled probe pairs (scheme B). Scheme B yields more information above a dimension-dependent threshold.
  \label{fig:fig3}}
 \end{figure}

Comparing just schemes O and B, Fig.~\ref{fig:fig3} shows that, above a $d$-dependent threshold for $\theta$, scheme B always yields more information (per channel use) than does scheme O and that below $\theta=1/\sqrt{3}$ scheme O always dominates scheme B.

\underline{Ancilla depolarization}: Identification scheme E uniformly dominates schemes O and B for the qudit depolarizing channel. Scheme E, though, may be unrealistic in assuming that the ancilla qudit in the entangled pair undergoes no degradation. Suppose we model the evolution of the ancilla qudit by a depolarizing channel $\Gamma_\eta$ with depolarizing probability $1-\eta$. If the probe and its ancilla qudit are initially in a maximally entangled state $\sigma=\ket{\mu}\bra{\mu}$ (see
Eq.~\eqref{eq:mu}), the channel output $\rho_\theta=(\Gamma_\eta\otimes\Gamma_\theta)(\sigma)$ is
\begin{equation}
\rho_\theta = \frac{1-\eta\theta}{d^2}\bI_{d^2} +\eta\theta\sigma.
\label{eq:enteta}
\end{equation}
This is identically the parametric family of states that would result from passing a $d^2$-dimensional qudit in a pure state
$\sigma$ with no external entanglement through a depolarizing channel with parameterization $h(\theta)=\eta\theta$. Therefore,
from (\ref{eq:qfi2}), the quantum Fisher information per use of $\Gamma_\theta$ with this scheme E$_\eta$ is
\begin{equation}
J_{\mathrm{E}_\eta}(\theta)
= \frac{\eta^2}{(1-\eta\theta)\left(\eta\theta+\frac{1}{d^2-1}\right)}.
\label{eq:infb2}
\end{equation}
Unlike identification scheme $\mathrm{E}=\mathrm{E}_1$, scheme $\mathrm{E}_\eta$ for $\eta<1$ does not uniformly dominate schemes O and B. In fact, for any degree of depolarization $\eta <1$ in the ancilla system, there exist $\theta$ where $J_{\mathrm{E}_\eta}(\theta) < J_{\mathrm{O}}(\theta)$ and/or $J_{\mathrm{E}_\eta}(\theta) < J_{\mathrm{B}}(\theta)$. Simple calculation shows, for example, that $J_{\mathrm{E}_\eta}(\theta) < J_{\mathrm{O}}(\theta)$ if $\theta > g(\eta)$ where
\begin{equation}
g(\eta) = \frac{\eta^2(d+1)-1}{\eta(1-\eta)(d^2-2)+\eta^2 d}.
\end{equation}
Since $g(\eta) <1$ for any $\eta<1$ and any $d$, there is always a$\theta$ large enough such that $J_{\mathrm{E}_\eta}(\theta) < J_{\mathrm{O}}(\theta)$. Similar statements can be made regarding $J_{\mathrm{E}_\eta}(\theta)$ relative to $J_{\mathrm{B}}(\theta)$. The broad point here is that, in this as in many other situations, realizing the theoretical advantage of entanglement depends on precise control of the experimental apparatus, and particularly in this case, evolution of the ancilla qudit.

\underline{Probe re-circulation}: One might ask whether any advantage can accrue to circulating the probe back through the channel, this being the scheme where the probe is passed through the channel, kept intact (no measurement), passed through the channel again, and, finally, measured. Conceivably, the probe could be re-circulated through the channel any number of times before measurement. Our schemes O, E, and B all accommodate probe re-circulation. The qudit depolarizing channel satisfies
$\Gamma_\theta^{\circ n} = \Gamma_{\theta^n}$ where $n$ is the number of channel passages so, from Eq.~\eqref{eq:qfi2},
\begin{subequations}
\begin{eqnarray}
J_{\mathrm{O,n}}(\theta) & = & \frac{n\theta^{2(n-1)}}
                                  {(1-\theta^n)
                                             \left(\theta^n+\frac{1}{d-1}
                                             \right)
                                        }
                                                         \\
J_{\mathrm{E,n}}(\theta) & = & \frac{n\theta^{2(n-1)}}
                                  {(1-\theta^n)
                                    \left(\theta^n+\frac{1}{d^2-1}
                                    \right)
                                  }
                             \\
J_{\mathrm{B,n}}(\theta) & = & \frac{2n\theta^{2(2n-1)}}
                                  {(1-\theta^{2n})
                                    \left(\theta^{2n}+\frac{1}{d^2-1}
                                    \right)
                                  }
\end{eqnarray}
\end{subequations}
where $J_{\mathrm{O,n}}(\theta)$, $J_{\mathrm{E,n}}(\theta)$ and $J_{\mathrm{B,n}}(\theta)$ are, respectively, the quantum Fisher
informations per channel use ($n$ channel uses with schemes O and E, $2n$ uses with B) about $\theta$ available by schemes O, E and B by measurement after $n$ circulations of the probe through the channel. Some elementary calculus shows that $J_{\mathrm{O,n+1}}(\theta)\leqslant J_{\mathrm{O,n}}(\theta)$ for all $n\geqslant 1$, $d\geqslant 2$ and $0\leqslant \theta\leqslant 1$, meaning that no degree of probe re-circulation is advantageous with scheme O. Since $J_{\mathrm{E,n}}(\theta;d) = J_{\mathrm{O,n}}(\theta;d^2)$ and
$J_{\mathrm{B,n}}(\theta;d) = J_{\mathrm{O,2n}}(\theta;d^2)$, the same must be true for schemes E and B. The information $J_{\mathrm{O,2}}(\theta)$ with scheme O,2 is shown in Fig.\ 2 for various channel dimensions $d$. We see that with increasing dimension the entanglement schemes E and B yield information increasingly close to schemes O,1 and O,2 without entanglement: $J_{\mathrm{E}}(\theta) \rightarrow J_{\mathrm{O,1}}(\theta)$ and $J_{\mathrm{B}}(\theta) \rightarrow J_{\mathrm{O,2}}(\theta)$.

We observed that probe re-circulation, with any of our schemes, can only diminish the quantum Fisher information in the channel output. These observations were made, though, on the basis of quantum Fisher information per channel use. If channel use is not a significant cost, probe re-circulation can be advantageous. In fact, when the channel depolarizing probability $1-\theta$ is small, re-circulating the probe just once can yield more information than entanglement. For the qubit depolarizing channel
with $\theta = .9$, for example, the quantum Fisher information gained by measuring a once re-circulated scheme O probe is $2J_{\mathrm{O,2}}(.9;2) = 9.42$, while that gained by measuring a maximally entangled probe with no re-circulation is $J_{\mathrm{E,1}}(.9;2) = 8.11$. Entanglement clearly holds promise for channel probing, but depending on the relative cost of this resource, probe re-circulation or some other similarly simple classical strategy may be more effective. In the case of the depolarizing
channel, since probe re-circulation may be, for example, just a matter of maintaining the probe in the target depolarizing medium for some longer period of time, entanglement may need to be very inexpensive to be competitive.

%%%%%%%%%%%%%%%%%%%%%%%%%%%%%%%%%%%%%%%%%%%%%%%%%%%%%%%%%%%%%%%%%%%%%%%%%%%%%%%%%%%%%%%%%%%%%%%%%%%%%%%%%
%%%%%%%%%%%%%%%%%%%                                                                       %%%%%%%%%%%%%%%
%%%%%%%%%%%%%%%%%%%         End section                                                 %%%%%%%%%%%%%%%
%%%%%%%%%%%%%%%%%%%                                                                       %%%%%%%%%%%%%%%
%%%%%%%%%%%%%%%%%%%%%%%%%%%%%%%%%%%%%%%%%%%%%%%%%%%%%%%%%%%%%%%%%%%%%%%%%%%%%%%%%%%%%%%%%%%%%%%%%%%%%%%%%

%%%%%%%%%%%%%%%%%%%%%%%%%%%%%%%%%%%%%%%%%%%%%%%%%%%%%%%%%%%%%%%%%%%%%%%%%%%%%%%%%%%%%%%%%%%%%%%%%%%%%%%%%
%%%%%%%%%%%%%%%%%%%                                                                       %%%%%%%%%%%%%%%
%%%%%%%%%%%%%%%%%%%         Begin section                                                 %%%%%%%%%%%%%%%
%%%%%%%%%%%%%%%%%%%                                                                       %%%%%%%%%%%%%%%
%%%%%%%%%%%%%%%%%%%%%%%%%%%%%%%%%%%%%%%%%%%%%%%%%%%%%%%%%%%%%%%%%%%%%%%%%%%%%%%%%%%%%%%%%%%%%%%%%%%%%%%%%

\section{Probing with partial entanglement}
\label{sec:partialentangle}

Various quantum Fisher informations $J(\theta)$ in the output of the depolarizing channel
probed by qudits with zero and maximal entanglement were presented and compared in the
previous section. Here we consider qudit probes just partially entangled with an ancilla.
The effect of partial entanglement is not obvious; channels are known for which a
partially entangled probe can yield more information than a fully entangled
probe~\cite{fujiwara04}. The essential conclusion of this section is that, for the qudit
depolarizing channel of any dimension, partial entanglement always yields information intermediate between $J_{\mathrm{O}}(\theta)$ from an unentangled qudit probe and $J_{\mathrm{E}}(\theta)$ from a maximally entangled probe.

The previous section's approach, represented by use of Eq.~\eqref{eq:qfi2}, is
applicable to certain special kinds of partial entanglement---qudits which are fully
entangled on just a subspace. We briefly pursue this point and then turn to the
question of partial entanglement generally. Consider a pair of qudits in the partially
entangled pure state $\sigma=\ket{\Psi}\bra{\Psi}$ with Schmidt decomposition
\begin{equation}
\ket{\Psi} =\psi_1 \ket{11}+\ldots+\psi_d\ket{dd}
\label{eq:schm}
\end{equation}
where $\psi_j \geqslant  0$ and $\psi_1^2+\dots+\psi_d^2=1$. Suppose that $d_o$ of the $\psi_i$ are $1/\sqrt{d_o}$ and that the remainder  are zero, where $2<d_o <d$. These qudits are maximally entangled on a subspace of dimension $d_o <d$ with no entanglement outside this subspace. The output of the depolarizing channel probed by the first of these qudits is
\begin{equation}
\rho_\theta = \frac{1-\theta}{dd_o}\bI_{d}\otimes\bI_{d_o}
+\theta\sigma
\end{equation}
and associated with this output is the quantum Fisher information
\begin{equation}
J(\theta)
=\frac{1}{(1-\theta)\left(\theta+\frac{1}{dd_o-1}\right)}.
\label{eq:pare}
\end{equation}
This shows for qudits with this special form of partial entanglement that
\begin{equation}
J_{\mathrm{O}}(\theta) \leqslant J(\theta) \leqslant J_{\mathrm{E}}(\theta).
\label{eq:sho}
\end{equation}
We now prove that Eq.~\eqref{eq:sho} is true generally for any partial entanglement.

Suppose two qudits are partially entangled generally as in Eq.~\eqref{eq:schm}. Probing the channel with the first qudit produces the channel output
\begin{equation}
\rho_\theta = (\Gamma_\theta\otimes\bI)(\sigma)
            =  \frac{1-\theta}{d}\; \bI_d \otimes \bD
               +\theta\sigma
\label{eq:out}
\end{equation}
where
\begin{equation}
  \bD = \sum_{i=1}^d \psi_i^2 \ket{i}\bra{i}
  \label{eq:Ddef}
\end{equation}
operates on a single qudit. Redefining $\bD := \sum_{i=1}^d \psi_i^2 \ket{ii}\bra{ii}$ as an operator on both qudits gives $\rho_\theta=\rho_C+\rho_Q$ where
\begin{equation}
 \rho_Q = \frac{1-\theta}{d}\bD
          +\theta \ket{\Psi}\bra{\Psi} \quad \textrm{and} \quad
 \rho_C = \frac{1-\theta}{d} \bC
\label{eq:rhos}
\end{equation}
are positive semi-definite operators with orthogonal supports $\cS = \{ \ket{ii}, i=1,\dots,d\}$ and $\cS^\perp=\{\ket{ij}, i\neq j\}$, respectively, determined by the Schmidt decomposition Eq.~\eqref{eq:schm}, and $\bC=\sum_{j\neq i} \psi_j^2 \ket{ij}\bra{ij}$.

A general result, whose proof is straightforward, is that, if $\rho_\theta = \rho_{1\theta} + \rho_{2\theta}$ where $\rho_{1\theta}$ and
$\rho_{2\theta}$ have orthogonal supports, then the score operator associated with $\rho_\theta$, i.e.\ $\bL$, can be expressed as $\bL = \bL_1 + \bL_2$ where $\bL_i$ is the score operator associated with $\rho_{i\theta}.$ Also the support of $\bL_i$ is orthogonal to that of $\bL_j$ and $\rho_{j\theta}$ if $i\neq j$. Thus $\bL=\bL_Q+\bL_C$ where $\bL_Q\rho_C=\bL_C\rho_Q=0$ such that $\bL_Q$ and $\bL_C$ satisfy, respectively,
\begin{eqnarray}
 \bL_Q\rho_Q+\rho_Q\bL_Q & = & 2\partial_\theta\rho_Q  = 2\left(\ket{\Psi}\bra{\Psi}-\frac{1}{d}\bD\right), 
                                               \label{eq:Q} \\
 \bL_C\rho_C+\rho_C\bL_C & = & 2\partial_\theta\rho_C  = -\frac{2}{d} \bC.
\label{eq:C}
\end{eqnarray}
We have immediately from Eq.~\eqref{eq:C}, then, that
\begin{equation}
 \bL_C= -\frac{1}{1-\theta}\, \bI_C
\end{equation}
 where $\bI_C$ is the identity on $\cS^\perp$. While no similarly simple solution exists for Eq.~\eqref{eq:Q}, substituting $\rho_Q$ from Eq.~\eqref{eq:rhos} in Eq.~\eqref{eq:Q} yields
\begin{equation}
\bra{\Psi}\bL_Q\ket{\Psi} = \frac{d-1}{d\theta}
                               -\frac{1-\theta}{d\theta}\Trace{(\bD\bL_Q)}.
\end{equation}

We now derive, using our results from Eqs.~\eqref{eq:Q} and~\eqref{eq:C}, an expression for the quantum Fisher information $J(\theta)$ in the channel output Eq.~\eqref{eq:out},
\begin{eqnarray}
J(\theta) & = & \Trace{(\bL \partial_\theta \rho_\theta)}
                \nonumber \\
          & = & \Trace{(\bL_C \partial_\theta \rho_C)}
              + \Trace{(\bL_Q \partial_\theta \rho_Q)}
                \nonumber \\
          & = & \Trace{\left[\bL_C \left( -\frac{1}{d}\bC \right)\right]}
              + \Trace{\left[\bL_Q \left( \ket{\Psi}\bra{\Psi}-\frac{1}{d}\bD \right) \right]}                                                              \nonumber \\
          & = & \frac{d-1}{d(1-\theta)} + \bra{\Psi}\bL_Q\ket{\Psi}
              - \frac{1}{d} \Trace{(\bD\bL_Q)}
                \nonumber  \\
          & = & \frac{d-1}{d(1-\theta)} + \frac{d-1}{d\theta}
               - \frac{1-\theta}{d\theta}\Trace{(\bD\bL_Q)} \nonumber \\
          &  & - \frac{1}{d} \Trace{(\bD\bL_Q)}
                 \nonumber  \\
          & = & \frac{d-1}{d\theta(1-\theta)}
               -\frac{1}{d\theta}\Trace{(\bD\bL_Q)}.
                \nonumber \\
          & = & \frac{d-1}{d\theta(1-\theta)}
                -\frac{1}{d\theta}\bl^{\tr} \bD\bx
                \label{eq:only}
\end{eqnarray}
where $\bx$ is the $d$ dimensional vector
\begin{equation}
   \bx = \begin{pmatrix}
            L_{11} \\
            L_{22} \\
            \vdots \\
            L_{dd}
           \end{pmatrix}
   \label{eq:xdef}
\end{equation}
with $L_{ii} = \bra{ii} \bL_Q \ket{ii}$, $\bD$ is, now, a $d\times d$ diagonal matrix with diagonal entries $\psi_1^2, \ldots , \psi_d^2$, and $\bl$ is the $d$ dimensional column vector of ones. Expression (\ref{eq:only}) shows, significantly, that only $\bL_Q$'s diagonal elements are needed for $J(\theta)$. In appendix~\ref{app:inversion} we demonstrate that
\begin{equation}
\bl^{\tr} \bD\bx = \frac{d-1}{1-\theta+d\theta}\,
                   - \frac{d^2\theta}{(1-\theta)(1-\theta+d\theta)^2}\,
                   \bb^{\tr}\bZ^{-1}\bb
\label{eq:med}
\end{equation}
where $\bb$ is a $d(d-1)/2$ dimensional vector with lexicographically ordered entries
\begin{equation}
   \bb = 2
           \begin{pmatrix}
            \psi_1 \psi_2\\
            \psi_1 \psi_3\\
            \vdots \\
            \psi_{d-1}\psi_d
           \end{pmatrix}
   \label{eq:bdef}
\end{equation}
and $\bZ$ is the $d(d-1)/2 \times d(d-1)/2$ matrix
\begin{equation}
    \bZ = \bJ + \alpha \bOmega^{\tr} \bOmega
\end{equation}
where $\alpha=d\theta/(1-\theta+d\theta).$ The two constituent matrices of $\bZ$ are
\begin{equation}
    \bJ = \begin{pmatrix}
           \psi_1^2+\psi_2^2 & 0 & \cdots & 0 \\
                 0 & \psi_1^2+\psi_3^2 & \cdots & 0 \\
                 \vdots & \vdots & \ddots  & \vdots \\
         0 & 0 & \cdots & \psi_{d-1}^2+\psi_d^2
          \end{pmatrix}
\end{equation}
and a $d \times d(d-1)/2$ matrix $\bOmega_d=f_d(\psi_1,\ldots,\psi_d)$ with $f_d(\cdot )$ is defined recursively by
\begin{subequations}
\begin{eqnarray}
 f_2(x,r)& = & \begin{pmatrix}
			            r \\ x
			         \end{pmatrix}
			         \quad
			         \textrm{and} \\
 f_{d+1}(x,\br_d) & = & 
         \begin{pmatrix}
           \br_d & \bz \\
           x\bI_d & f_d(\br_d)
         \end{pmatrix}
    \label{eq:omegadef}
\end{eqnarray}
\end{subequations}
where $\br_d=\left( r_1 \ldots r_d \right)$ and $\bz$ is a row vector of $(d-1)(d-2)/2$ zeros. For example
\begin{eqnarray}
\bOmega_4 & = & f_4(\psi_1,\psi_2,\psi_3,\psi_4)
                \nonumber \\
                    & = & \begin{pmatrix}
                              \psi_2 & \psi_3 & \psi_4 &0 &0 &0 \\
                                        \psi_1 & 0 & 0 & \psi_3 & \psi_4 & 0 \\
                                        0 & \psi_1 & 0 & \psi_2 & 0 & \psi_4 \\
                                        0 & 0 & \psi_1 & 0 & \psi_2 &  \psi_3
                \end{pmatrix}.
\end{eqnarray}
Combining Eqs.~\eqref{eq:only} and~\eqref{eq:med} yields
\begin{eqnarray}
	 J(\theta) & = & \frac{1}{(1-\theta)(\theta+\frac{1}{d-1})}\nonumber \\
	           &  & + \frac{d}{(1-\theta)(1-\theta+d\theta)^2} \bb^{\tr}\bZ^{-1} \bb.
\label{eq:parent}
\end{eqnarray}
While $\bZ$ is singular in cases where the qudits' entanglement is confined to a proper subspace, $\bb^{\tr}\bZ^{-1} \bb$ is always well-defined. With this in mind one can readily check that, for maximal entanglement on a subspace, Eq.~\eqref{eq:parent} is identically Eq.~\eqref{eq:pare}.

To prove Eq.~\eqref{eq:sho} we must prove that
\begin{equation}
\bb^{\tr}\bZ^{-1} \bb \leqslant
(d-1)\frac{1-\theta+d\theta}{1-\theta+d^2\theta}
\end{equation}
or, equivalently, that
\begin{equation}
\bb^{\tr}(\alpha\bO^{\tr}\bO+\bJ)^{-1} \bb
\leqslant \frac{1}{\alpha+\frac{1}{d-1}}  \; .
\label{eq:ineq2}
\end{equation}
Let $\lambda_{\mathrm{min}}[\bM]$ and $\lambda_{\mathrm{max}}[\bM]$ denote, respectively, the minimum and the maximum eigenvalues of a positive semi-definite matrix $\bM$, and let $\bA_d$ denote the $d\times d$ matrix $\bA_d=\bO \bJ^{-1}\bO^{\tr}$. Let $\bpsi$ be the length-$d$ column vector with components $\psi_j$. Using the relation $\bb=\bO^{\tr} \bpsi$, the matrix inversion lemma \cite{horn85}, and the Rayleigh-Ritz theorem \cite{horn85}, we have
\begin{eqnarray}
\bb^{\tr}(\alpha\bO^{\tr}\bO+\mathbf{J})^{-1} \bb & = & \bpsi^{\tr}\bO(\alpha\bO^{\tr}\bO+\mathbf{J})^{-1} \bO^{\tr} \bpsi
                                                        \nonumber \\
                                                  & = & \bpsi^{\tr} \left(\alpha\bI_d+\bA_d^{-1} \right)^{-1} \bpsi
                                                        \nonumber \\
                                                  & \leqslant & \lambda_{\max}
                                                           \left[
                                                            \left( \alpha\bI_d+\bA_d^{-1}
                                                            \right)^{-1}
                                                           \right]
                                                           \max_{\bpsi}\left( \bpsi^{\tr}\bpsi\right)  \nonumber \\
                                                  & = & \lambda_{\min}^{-1}
                                                        \left[
                                                          \alpha\bI_d+\bA_d^{-1}
                                                        \right]
                                                        \nonumber \\
                                                  & = & \frac{1}{\alpha+\lambda_{\min} [\bA_d^{-1}]}
                                                        \nonumber \\
                                                  & = & \frac{1}{\alpha+\lambda_{\max}^{-1}[\bA_d]}.
                                                        \label{eq:fin}
\end{eqnarray}
According to Eqs.~\eqref{eq:ineq2} and~\eqref{eq:fin}, we only need to show that $\lambda_{\max}[\bA_d]=d-1$ to prove the inequality of Eq.~\eqref{eq:sho}. First we show that $d-1$ is an eigenvalue of $\bA_d$; then we show that it is the maximum eigenvalue. Let $\bLambda$ be an eigenvector of $\bA_d$ with associated eigenvalue $\lambda$ so that
\begin{equation}
\bA_d \bLambda = \lambda\bLambda.
\end{equation}
Let $\bLambda_1= \bJ^{-1}\bO^{\tr} \bLambda$. Then
\begin{equation}
\bJ^{-1}\bO^{\tr}\bO \bLambda_1 = \lambda\bLambda_1  \; ,
\label{eq:eig}
\end{equation}
so $\lambda$ is also an eigenvalue of $\bJ^{-1}\bO^{\tr}\bO$. One can check by direct substitution that Eq.~\eqref{eq:eig} is satisfied by $\lambda = d-1$ with (unnormalized) eigenvector
\begin{equation}
\bLambda_1 = \begin{pmatrix}
              \dfrac{1}{\psi_1\psi_2} &
              \dfrac{1}{\psi_1\psi_3} &
              \cdots &
              \dfrac{1}{\psi_{d-1}\psi_d}
             \end{pmatrix}^{\tr}.
\end{equation}
In appendix~\ref{app:maxeigenvalue} we prove that $d-1$ is the maximum eigenvalue of $\bA_d$, and this completes the proof of Eq.~\eqref{eq:sho}.

%%%%%%%%%%%%%%%%%%%%%%%%%%%%%%%%%%%%%%%%%%%%%%%%%%%%%%%%%%%%%%%%%%%%%%%%%%%%%%%%%%%%%%%%%%%%%%%%%%%%%%%%%
%%%%%%%%%%%%%%%%%%%                                                                       %%%%%%%%%%%%%%%
%%%%%%%%%%%%%%%%%%%         End section                                                 %%%%%%%%%%%%%%%
%%%%%%%%%%%%%%%%%%%                                                                       %%%%%%%%%%%%%%%
%%%%%%%%%%%%%%%%%%%%%%%%%%%%%%%%%%%%%%%%%%%%%%%%%%%%%%%%%%%%%%%%%%%%%%%%%%%%%%%%%%%%%%%%%%%%%%%%%%%%%%%%%

%%%%%%%%%%%%%%%%%%%%%%%%%%%%%%%%%%%%%%%%%%%%%%%%%%%%%%%%%%%%%%%%%%%%%%%%%%%%%%%%%%%%%%%%%%%%%%%%%%%%%%%%%
%%%%%%%%%%%%%%%%%%%                                                                       %%%%%%%%%%%%%%%
%%%%%%%%%%%%%%%%%%%         Begin section                                                 %%%%%%%%%%%%%%%
%%%%%%%%%%%%%%%%%%%                                                                       %%%%%%%%%%%%%%%
%%%%%%%%%%%%%%%%%%%%%%%%%%%%%%%%%%%%%%%%%%%%%%%%%%%%%%%%%%%%%%%%%%%%%%%%%%%%%%%%%%%%%%%%%%%%%%%%%%%%%%%%%

\section{Optimal measurements}
\label{sec:measurements}

Realizing the precision promised by the quantum Fisher information in the Cram\'{e}r-Rao inequality depends on optimal measurement of the channel output. The quantum measurement that saturates the inequality is not always obvious, and in some cases---certain vectors of parameters, for example---no such measurement exists \cite{chen07}. For the qudit depolarizing channel, is is possible to describe a measurement which yields a (classical) Fisher information which equals the quantum Fisher information. For this channel the density operator after evolution is  such that its eigenstates are independent of the channel parameter (only the eigenvalues depend on the channel parameter), i.e.\
\begin{equation}
    \rho_\theta = \sum_i \lambda_i(\theta) \ket{s_i}\bra{s_i}
\end{equation}
where $\lambda_i(\theta)$ and $\ket{s_i}$ are the eigenvalues and eigenstates respectively. Then~\cite{paris09} the quantum Fisher information is
\begin{equation}
    J(\theta) = \sum_i \frac{1}{\lambda_i(\theta)}\; \left( \frac{d \lambda_i}{d \theta} \right)^2.
\end{equation}
However~\cite{paris09}, this is exactly the expression for the classical Fisher information for a probability distribution with probabilities $\left\{ \lambda_i(\theta)\right\}.$ Thus the quantum Fisher information can be attained via a measurement which results in this probability distribution. Clearly measurement in the eigenbasis of $\rho_\theta$ accomplishes this.  Here, one of the eigenstates of $\rho_\theta$ is the initial state $\ket{\mu}$ of the pair of qudits prior to action by the depolarizing channel. It follows that the projective measurement for which the projectors are
\begin{align}
  \Pi_0 & =  \ket{\mu}\bra{\mu} \quad \textrm{and}\\
  \Pi_1 & =  I - \ket{\mu}\bra{\mu}
\end{align}
yields a probability distribution, which gives a classical Fisher information exactly equal to that of the quantum Fisher information. For an entangled input state such as that used in scheme E, this optimal measurement requires measurements involving entangled states. It is unclear whether there is a scheme which involves unentangled states at the measurement phase and which attains the quantum Fisher information; such a scheme exists for unitary parameter estimation~\cite{giovannetti06}.

%%%%%%%%%%%%%%%%%%%%%%%%%%%%%%%%%%%%%%%%%%%%%%%%%%%%%%%%%%%%%%%%%%%%%%%%%%%%%%%%%%%%%%%%%%%%%%%%%%%%%%%%%
%%%%%%%%%%%%%%%%%%%                                                                       %%%%%%%%%%%%%%%
%%%%%%%%%%%%%%%%%%%         Begin section                                                 %%%%%%%%%%%%%%%
%%%%%%%%%%%%%%%%%%%                                                                       %%%%%%%%%%%%%%%
%%%%%%%%%%%%%%%%%%%%%%%%%%%%%%%%%%%%%%%%%%%%%%%%%%%%%%%%%%%%%%%%%%%%%%%%%%%%%%%%%%%%%%%%%%%%%%%%%%%%%%%%%

\section{Summary}
\label{sec:summary}

We have given an extensive treatment of channel identification for the general qudit depolarizing channel, comparing different probe preparation schemes based on the quantum Fisher information attainable by channel probing. We showed for depolarizing channels of any dimension that maximally entangling the channel probe with an external qubit uniformly realizes the maximum theoretical advantage, but that this advantage disappears if the ancilla qubit is not sufficiently shielded from depolarization. We
primarily compared the same three channel identification schemes considered in \cite{fujiwara01, sasaki02}, though our results apply generally for dimension $d <\infty$. We departed from \cite{fujiwara01, sasaki02} most significantly in choosing to compare identification schemes on an information per channel use basis. This is a practically meaningful basis of comparison; in any case, our results for each scheme are readily amenable to other comparisons.

We have considered here identification schemes involving bipartite probe entanglement, with an ancilla qudit or with another probe. Multipartite probe entanglement of various orders and types remains for study. We hope to be able to address at least some important multipartite entanglement schemes with our present methods.

%%%%%%%%%%%%%%%%%%%%%%%%%%%%%%%%%%%%%%%%%%%%%%%%%%%%%%%%%%%%%%%%%%%%%%%%%%%%%%%%%%%%%%%%%%%%%%%%%%%%%%%%%
%%%%%%%%%%%%%%%%%%%                                                                       %%%%%%%%%%%%%%%
%%%%%%%%%%%%%%%%%%%         End section                                                 %%%%%%%%%%%%%%%
%%%%%%%%%%%%%%%%%%%                                                                       %%%%%%%%%%%%%%%
%%%%%%%%%%%%%%%%%%%%%%%%%%%%%%%%%%%%%%%%%%%%%%%%%%%%%%%%%%%%%%%%%%%%%%%%%%%%%%%%%%%%%%%%%%%%%%%%%%%%%%%%%

\appendix

%%%%%%%%%%%%%%%%%%%%%%%%%%%%%%%%%%%%%%%%%%%%%%%%%%%%%%%%%%%%%%%%%%%%%%%%%%%%%%%%%%%%%%%%%%%%%%%%%%%%%%%%%
%%%%%%%%%%%%%%%%%%%                                                                       %%%%%%%%%%%%%%%
%%%%%%%%%%%%%%%%%%%         Begin section                                                 %%%%%%%%%%%%%%%
%%%%%%%%%%%%%%%%%%%                                                                       %%%%%%%%%%%%%%%
%%%%%%%%%%%%%%%%%%%%%%%%%%%%%%%%%%%%%%%%%%%%%%%%%%%%%%%%%%%%%%%%%%%%%%%%%%%%%%%%%%%%%%%%%%%%%%%%%%%%%%%%%

\section{Diagonal elements of the score operator}
\label{app:inversion}

Substituting from Eq.~\eqref{eq:rhos} into Eq.~\eqref{eq:Q} gives
\begin{eqnarray}
    -\frac{1}{d}\bD + \ket{\Psi}\bra{\Psi} &  = & \frac{1-\theta}{2d}\,
                                             \left(
                                              \bL_Q \bD + \bD\bL_Q
                                             \right) \nonumber \\
                                           & &  +
                                             \frac{\theta}{2}
                                             \left(
                                              \bL_Q \ket{\Psi}\bra{\Psi} + \ket{\Psi}\bra{\Psi} \bL_Q
                                             \right).
\end{eqnarray}
Acting with this on $\ket{jj}$ and acting on the result with $\bra{ii}$ gives
\begin{eqnarray}
    -\frac{2}{d}\psi_i^2 \delta_{ij} + 2\psi_i \psi_j  &  = & \frac{1-\theta}{d}\,
                                             \left( \psi_i^2 + \psi_j^2 \right) L_{ij} \nonumber \\
                                        &  \!\!\!\!\! + & \!\!\!
                                             \theta
                                             \sum_{k=1}^d
                                             \left(
                                               \psi_j \psi_k L_{ik} + \psi_i \psi_k L_{jk}
                                             \right).
    \label{eq:linearforL}
\end{eqnarray}
where $L_{ij} = \bra{ii} \bL_Q \ket{jj}.$ Note that the score operator is Hermitian and in this case real; thus $L_{ij} = L_{ji}.$ Thus Eq.~\eqref{eq:linearforL} is a system of $d(d+1)/2$ distinct linear equations for the components of $\bL_Q.$ Our aim is to invert these for the diagonal elements of $\bL_Q.$ Using $\bx$ as defined in Eq.~\eqref{eq:xdef} and
\begin{equation}
    \by = \begin{pmatrix}
            L_{12} \\
            L_{13} \\
            \vdots \\
            L_{1d} \\
            L_{23} \\
            \vdots \\
            L_{d-1 d}
           \end{pmatrix}
\end{equation}
gives
\begin{equation}
    \begin{pmatrix}
     \bR & \bS \\ \bS^{\tr} & \bT
    \end{pmatrix}
    \begin{pmatrix}
     \bx \\
     \by
    \end{pmatrix}
    =
    \begin{pmatrix}
     \ba \\
     \bb
    \end{pmatrix}
    \label{eq:mainlineareq}
\end{equation}
where $\ba$ is a $d$ dimensional vector given by
\begin{equation}
    \ba  = \frac{d-1}{d}
            \begin{pmatrix}
              \psi_1^2 \\
              \psi_2^2 \\
              \vdots \\
              \psi_d^2
            \end{pmatrix},
\end{equation}
$\bb$ is given in Eq.~\eqref{eq:bdef}, and the concatenation $\begin{pmatrix} \bx & \by \end{pmatrix}^{\tr}$ is the column vector consisting of the components of $\bx$ followed by those of $\by.$ The submatrices $\bR$ and $\bS$ within Eq.~\eqref{eq:mainlineareq} are readily derived from Eq.~\eqref{eq:linearforL} and are
\begin{equation}
    \bR = \begin{pmatrix}
                    \rho_{11} & 0 & \cdots & 0 \\
                    0 & \rho_{22} & \cdots & 0 \\
                  \vdots & \vdots & \ddots & \vdots \\
          0 & 0 & \cdots & \rho_{dd}
         \end{pmatrix}
       = \frac{1-\theta+d\theta}{d} \bD
\end{equation}
and
\begin{equation}
    \bS = \theta \bD^{1/2} \bOmega_d
    \label{eq:Sdef}
\end{equation}
where $\bOmega_d$ is as defined via Eq.~\eqref{eq:omegadef}. The final submatrix is
\begin{equation}
    \bT=\frac{1-\theta}{d} \bJ +\theta\bOmega_d^{\tr}\bOmega_d
    \label{eq:Tdef}
\end{equation}
and this can be verified by considering Eq.~\eqref{eq:linearforL} for $i < j.$ The left side yields the row of $\bb$ corresponding to the indices $i$ and $j$ and the right side yields the following contribution from $\bT \by$
\begin{eqnarray}
  & & \frac{1-\theta}{d}\,
        \left( \psi_i^2 + \psi_j^2 \right) L_{ij} \nonumber \\
  & & 
          + \theta
           \left(
           \sum_{k=1, k\neq i}^d \psi_j \psi_k L_{ik}
            + \sum_{k=1, k\neq j}^d\psi_i \psi_k L_{jk}
           \right).
\end{eqnarray}
The first term clearly corresponds to $(1-\theta)/d \; \; \bJ \by.$ To check the second term, we can show by an iterative process that
\begin{equation}
    \bOmega_d \by = \begin{pmatrix}
                     \sum_{k\neq 1} \psi_k L_{1k} \\
                     \vdots  \\
                     \sum_{k\neq d} \psi_k L_{dk}
                    \end{pmatrix}.
\end{equation}
Using this, we can again show iteratively that
\begin{equation}
    \bOmega_d^{\tr}\bOmega_d \by
                 = \begin{pmatrix}
                     \sum_{k\neq 1} \psi_k L_{1k} + \sum_{k\neq 2} \psi_k L_{2k} \\
                     \sum_{k\neq 1} \psi_k L_{1k} + \sum_{k\neq 3} \psi_k L_{3k} \\
                     \vdots  \\
                     \sum_{k\neq d-1} \psi_k L_{d-1k} + \sum_{k\neq d} \psi_k L_{dk}
                    \end{pmatrix}
\end{equation}
and this corresponds to the second term. This justifies the form of Eq.~\eqref{eq:mainlineareq}.

Applying the matrix inversion lemma~\cite{horn85} to the solution of Eq.~\eqref{eq:mainlineareq} for $\bx$ gives
\begin{eqnarray}
\bx & = & \left( \bR-\bS\bT^{-1}\bS^{\tr} \right)^{-1}
          \left( \ba-\bS\bT^{-1}\bb \right) \nonumber \\
    & = & \left[ \bR^{-1} +\bR^{-1}\bS(\bT-\bS^{\tr} \bR^{-1}\bS)^{-1}
                 \bS^{\tr}\bR^{-1}
          \right]   \nonumber \\
    &  &   (\ba-\bS\bT^{-1}\bb)
\label{eq:x}
\end{eqnarray}
From (\ref{eq:x}), and using
\begin{equation}
\bR^{-1}\ba = \frac{d-1}{1-\theta+d\theta} \bl
\end{equation}
a lengthy calculation yields
\begin{eqnarray}
\bD\bx & = & \frac{d}{1-\theta+d\theta} ( \ba-\bS\bT^{-1}\bb ) \nonumber \\
       & + & \frac{d}{1-\theta+d\theta} \bS(\bT-\bS^{\tr}\bR^{-1}\bS)^{-1}\bS^{\tr} \nonumber \\
       &  & 
           \left[ \frac{d-1}{1-\theta+d\theta}\bl -\bR^{-1}\bS\bT^{-1}\bb
           \right] .
\label{eq:Dx}
\end{eqnarray}
Using the relations $\bl^{\tr}\ba=(d-1)/d$ and $\bS^{\tr} \bl = \theta\bb$, we calculate from Eq.~\eqref{eq:Dx} that
\begin{eqnarray}
\bl^{\tr} \bD\bx & = & \frac{d-1}{1-\theta+d\theta} + \frac{d\theta}{(1-\theta+d\theta)^2} \nonumber \\
                 &  & \times
                    \bb^{\tr}(\bS^{\tr}\bR^{-1}\bS-\bT)^{-1}\bb.
\end{eqnarray}
Recalling the definitions of $\bS$ and $\bT$ in Eqs.~\eqref{eq:Sdef} and~\eqref{eq:Tdef}, we calculate further that
\begin{eqnarray}
\bl^{\tr} \bD\bx & = &\frac{d-1}{1-\theta+d\theta} \nonumber \\
                 & & - \frac{d^2\theta}{(1-\theta)(1-\theta+d\theta)^2}
                   \bb^{\tr}\bZ^{-1}\bb
\label{eq:medapp}
\end{eqnarray}
where
\begin{equation}
\bZ =\bJ+  \alpha\bOmega^{\tr} \bOmega
\end{equation}
with $\alpha=d\theta/(1-\theta+d\theta)$. This gives Eq.~\eqref{eq:med}.

%%%%%%%%%%%%%%%%%%%%%%%%%%%%%%%%%%%%%%%%%%%%%%%%%%%%%%%%%%%%%%%%%%%%%%%%%%%%%%%%%%%%%%%%%%%%%%%%%%%%%%%%%
%%%%%%%%%%%%%%%%%%%                                                                       %%%%%%%%%%%%%%%
%%%%%%%%%%%%%%%%%%%         End section                                                 %%%%%%%%%%%%%%%
%%%%%%%%%%%%%%%%%%%                                                                       %%%%%%%%%%%%%%%
%%%%%%%%%%%%%%%%%%%%%%%%%%%%%%%%%%%%%%%%%%%%%%%%%%%%%%%%%%%%%%%%%%%%%%%%%%%%%%%%%%%%%%%%%%%%%%%%%%%%%%%%%

%%%%%%%%%%%%%%%%%%%%%%%%%%%%%%%%%%%%%%%%%%%%%%%%%%%%%%%%%%%%%%%%%%%%%%%%%%%%%%%%%%%%%%%%%%%%%%%%%%%%%%%%%
%%%%%%%%%%%%%%%%%%%                                                                       %%%%%%%%%%%%%%%
%%%%%%%%%%%%%%%%%%%         Begin section                                                 %%%%%%%%%%%%%%%
%%%%%%%%%%%%%%%%%%%                                                                       %%%%%%%%%%%%%%%
%%%%%%%%%%%%%%%%%%%%%%%%%%%%%%%%%%%%%%%%%%%%%%%%%%%%%%%%%%%%%%%%%%%%%%%%%%%%%%%%%%%%%%%%%%%%%%%%%%%%%%%%%

\section{Maximum eigenvalue of $\bA_d$}
\label{app:maxeigenvalue}

The matrix $\bA_d=[a_{ij}]$ has elements
\begin{equation}
a_{ij}=\begin{cases}
             \sum_{k=1,k\ne i}^d
             \dfrac{\psi_k^2}{\psi_i^2+\psi_k^2}  & \textrm{if $i=j$,} \\
             \dfrac{\psi_i\psi_j}{\psi_i^2+\psi_j^2} & \textrm{if $i\neq j$.}
       \end{cases}
\label{eq:a}
\end{equation}
\begin{widetext}
For example,
\begin{equation}
\bA_3 = \begin{pmatrix}
          \frac{\psi_2^2}{\psi_1^2+\psi_2^2}+\frac{\psi_3^2}{\psi_1^2+\psi_3^2} &
                    \frac{\psi_1\psi_2}{\psi_1^2+\psi_2^2} &
                    \frac{\psi_1\psi_3}{\psi_1^2+\psi_3^2} \\
                    \frac{\psi_2\psi_1}{\psi_1^2+\psi_2^2} &
                    \frac{\psi_1^2}{\psi_2^2+\psi_1^2}+\frac{\psi_3^2}{\psi_2^2+\psi_3^2} &
                    \frac{\psi_2\psi_3}{\psi_2^2+\psi_3^2} \\
                    \frac{\psi_1\psi_3}{\psi_1^2+\psi_3^2} &
                    \frac{\psi_2\psi_3}{\psi_2^2+\psi_3^2} &
                    \frac{\psi_1^2}{\psi_3^2+\psi_1^2}+\frac{\psi_2^2}{\psi_3^2+\psi_2^2}
        \end{pmatrix}.
\end{equation}
\end{widetext}
Define the angles $\phi_{ij}$ for $1\leqslant i<j\leqslant d$ by $ \cos\phi_{ij} =  \psi_i/\sqrt{\psi_i^2+\psi_j^2}$ and let $\Phi_d$ be the length-$d(d-1)/2$ vector $\Phi_d = (\phi_{12} \;\; \phi_{13} \;\ldots\; \phi_{d-1,d})$.
The matrix $\bA_d$ is a function $\bA_d(\Phi_d)$ of the angles in $\Phi_d$. Consider the characteristic function $h(\lambda)=\det (\lambda \bI_d-\bA_d)$ of $\bA_d$. The eigenvalue $\lambda_{\max}[\bA_d] = d-1$ if and only if $h^\prime(\lambda)$ is positive for $\lambda>d-1$. 
This is a consequence of the following. If the eigenvalues of $\bA_d$ are $\lambda_\mathrm{min}, \ldots, \lambda_{\max}$ then $h(\lambda) = (\lambda - \lambda_\mathrm{min}) \ldots (\lambda - \lambda_{\max})$ and $h^\prime(\lambda) = h(\lambda)\left[ 1/(\lambda - \lambda_\mathrm{min}) + \ldots 1/(\lambda - \lambda_{\max})\right].$ Clearly if $\lambda > \lambda_{\max}$ then $h(\lambda)>0$ and $h^\prime(\lambda)>0$. A graphical representation of the characteristic function reveals that at some point between $\lambda_{\max}$ and the eigenvalue immediately smaller than this, $h^\prime(\lambda)$ must be negative since $h(\lambda)$ in this region is between two of its roots. Thus if $h^\prime(\lambda)>0$ for all $\lambda > \tilde{\lambda}$ then $\tilde{\lambda} = \lambda_{\max}.$
We have
\begin{equation}
h^\prime(\lambda) = \sum_{i=1}^d\det(\lambda\bI_{d-1}-\bA_{d}^{ii})
\label{eq:der}
\end{equation}
where $\bA_d^{ii}$ is the matrix $\bA_d$ with its $i$th row and column deleted. The summand in Eq.~\eqref{eq:der} is the characteristic function of $\bA_d^{ii}$, so we need just show that no $\bA_d^{ii}$ has an eigenvalue greater than $d-1$. A standard calculation yields %
\begin{equation}
\bA_d^{ii}=A_{d-1}(\Phi_d^i) + \Delta^i
\end{equation}
where $\Delta^i$ is a diagonal matrix with diagonal elements $\cos^2\phi_{ij}$, and $\Phi_d^i$ is the vector $\Phi_d$ with the $d-1$ angles $\phi_{ij}$ and $\phi_{ji}$ deleted. Therefore, by Weyl's inequality \cite{horn85},
\begin{eqnarray}
\lambda_{\max}[\bA_d^{ii}] & \leqslant & \lambda_{\max}[A_{d-1}(\Phi_d^i)] + \lambda_{\max}[\Delta^i]
                                         \nonumber \\
                           & \leqslant & \lambda_{\max}[A_{d-1}(\Phi_d^i)] + 1.
\label{eq:bd}
\end{eqnarray}
We have $\lambda_{\max}[\bA_2(\Phi_2)] =\lambda_{\max}[\bA_2(\Phi_3^{i})]=1$ by direct calculation from Eq.~\eqref{eq:a}. Thus $\lambda_{\max}[\bA_3^{ii}]\leqslant 2$ by Eq.~\eqref{eq:bd}, and $\lambda_{\max}[\bA_3]\leqslant 2$ by consideration of Eq.~\eqref{eq:der}.
Arguing recursively, we conclude that $\lambda_{\max}[\bA_d]\leqslant d-1$ for $d \geqslant 2$. We saw that $d-1$ is an eigenvalue of $\bA_d$ so $\lambda_{\max}[\bA_d]= d-1$.
 
%%%%%%%%%%%%%%%%%%%%%%%%%%%%%%%%%%%%%%%%%%%%%%%%%%%%%%%%%%%%%%%%%%%%%%%%%%%%%%%%%%%%%%%%%%%%%%%%%%%%%%%%%
%%%%%%%%%%%%%%%%%%%                                                                       %%%%%%%%%%%%%%%
%%%%%%%%%%%%%%%%%%%         End section                                                 %%%%%%%%%%%%%%%
%%%%%%%%%%%%%%%%%%%                                                                       %%%%%%%%%%%%%%%
%%%%%%%%%%%%%%%%%%%%%%%%%%%%%%%%%%%%%%%%%%%%%%%%%%%%%%%%%%%%%%%%%%%%%%%%%%%%%%%%%%%%%%%%%%%%%%%%%%%%%%%%%

%\bibliography{../biblio/refs}
%\bibliography{refs}

\begin{thebibliography}{10}%
\makeatletter
\providecommand \@ifxundefined [1]{%
 \ifx #1\undefined \expandafter \@firstoftwo
 \else \expandafter \@secondoftwo
\fi
}%
\providecommand \@ifnum [1]{%
 \ifnum #1\expandafter \@firstoftwo
 \else \expandafter \@secondoftwo
\fi
}%
\providecommand \enquote [1]{``#1''}%
\providecommand \bibnamefont  [1]{#1}%
\providecommand \bibfnamefont [1]{#1}%
\providecommand \citenamefont [1]{#1}%
\providecommand\href[0]{\@sanitize\@href}%
\providecommand\@href[1]{\endgroup\@@startlink{#1}\endgroup\@@href}%
\providecommand\@@href[1]{#1\@@endlink}%
\providecommand \@sanitize [0]{\begingroup\catcode`\&12\catcode`\#12\relax}%
\@ifxundefined \pdfoutput {\@firstoftwo}{%
 \@ifnum{\z@=\pdfoutput}{\@firstoftwo}{\@secondoftwo}%
}{%
 \providecommand\@@startlink[1]{\leavevmode}%
 \providecommand\@@endlink[0]{}%
}{%
 \providecommand\@@startlink[1]{%
  \leavevmode
  \pdfstartlink
   attr{/Border[0 0 1 ]/H/I/C[0 1 1]}%
   user{/Subtype/Link/A<</Type/Action/S/URI/URI(#1)>>}%
  \relax
 }%
 \providecommand\@@endlink[0]{\pdfendlink}%
}%
\providecommand \url  [0]{\begingroup\@sanitize \@url }%
\providecommand \@url [1]{\endgroup\@href {#1}{\urlprefix}}%
\providecommand \urlprefix [0]{URL }%
\providecommand \Eprint[0]{\href }%
\@ifxundefined \urlstyle {%
  \providecommand \doi [1]{doi:\discretionary{}{}{}#1}%
}{%
  \providecommand \doi [0]{doi:\discretionary{}{}{}\begingroup
  \urlstyle{rm}\Url }%
}%
\providecommand \doibase [0]{http://dx.doi.org/}%
\providecommand \Doi[1]{\href{\doibase#1}}%
\providecommand \bibAnnote [3]{%
  \BibitemShut{#1}%
  \begin{quotation}\noindent
    \textsc{Key:}\ #2\\\textsc{Annotation:}\ #3%
  \end{quotation}%
}%
\providecommand \bibAnnoteFile [2]{%
  \IfFileExists{#2}{\bibAnnote {#1} {#2} {\input{#2}}}{}%
}%
\providecommand \typeout [0]{\immediate \write \m@ne }%
\providecommand \selectlanguage [0]{\@gobble}%
\providecommand \bibinfo [0]{\@secondoftwo}%
\providecommand \bibfield [0]{\@secondoftwo}%
\providecommand \translation [1]{[#1]}%
\providecommand \BibitemOpen[0]{}%
\providecommand \bibitemStop [0]{}%
\providecommand \bibitemNoStop [0]{.\EOS\space}%
\providecommand \EOS [0]{\spacefactor3000\relax}%
\providecommand \BibitemShut [1]{\csname bibitem#1\endcsname}%
%</preamble>
\bibitem{fujiwara01}%
  \BibitemOpen
  \bibfield{author}{%
  \bibinfo {author} {\bibfnamefont{A.}~\bibnamefont{Fujiwara}},\ }%
  \bibfield{journal}{%
  \Doi{10.1103/PhysRevA.63.042304}{\bibinfo {journal} {Phys. Rev. A}}\ }%
  \textbf{\bibinfo {volume} {63}},\ \bibinfo {pages} {042304} (\bibinfo {month}
  {Mar}\ \bibinfo {year} {2001})%
  \bibAnnoteFile{NoStop}{fujiwara01}%
\bibitem{sasaki02}%
  \BibitemOpen
  \bibfield{author}{%
  \bibinfo {author} {\bibfnamefont{M.}~\bibnamefont{Sasaki}}, \bibinfo {author}
  {\bibfnamefont{M.}~\bibnamefont{Ban}},\ and\ \bibinfo {author}
  {\bibfnamefont{S.~M.}\ \bibnamefont{Barnett}},\ }%
  \bibfield{journal}{%
  \Doi{10.1103/PhysRevA.66.022308}{\bibinfo {journal} {Phys. Rev. A}}\ }%
  \textbf{\bibinfo {volume} {66}},\ \bibinfo {pages} {022308} (\bibinfo {month}
  {Aug}\ \bibinfo {year} {2002})%
  \bibAnnoteFile{NoStop}{sasaki02}%
\bibitem{fujiwara03}%
  \BibitemOpen
  \bibfield{author}{%
  \bibinfo {author} {\bibfnamefont{A.}~\bibnamefont{Fujiwara}}\ and\ \bibinfo
  {author} {\bibfnamefont{H.}~\bibnamefont{Imai}},\ }%
  \bibfield{journal}{%
  \bibinfo {journal} {J. Phys. A}\ }%
  \textbf{\bibinfo {volume} {36}},\ \bibinfo {pages} {8093}%
  \bibAnnoteFile{NoStop}{fujiwara03}%
\bibitem{fujiwara04}%
  \BibitemOpen
  \bibfield{author}{%
  \bibinfo {author} {\bibfnamefont{A.}~\bibnamefont{Fujiwara}},\ }%
  \bibfield{journal}{%
  \Doi{10.1103/PhysRevA.70.012317}{\bibinfo {journal} {Phys. Rev. A}}\ }%
  \textbf{\bibinfo {volume} {70}},\ \bibinfo {pages} {012317} (\bibinfo {month}
  {Jul}\ \bibinfo {year} {2004})%
  \bibAnnoteFile{NoStop}{fujiwara04}%
\bibitem{ballester04}%
  \BibitemOpen
  \bibfield{author}{%
  \bibinfo {author} {\bibfnamefont{M.~A.}\ \bibnamefont{Ballester}},\ }%
  \bibfield{journal}{%
  \Doi{10.1103/PhysRevA.69.022303}{\bibinfo {journal} {Phys. Rev. A}}\ }%
  \textbf{\bibinfo {volume} {69}},\ \bibinfo {pages} {022303} (\bibinfo {month}
  {Feb}\ \bibinfo {year} {2004})%
  \bibAnnoteFile{NoStop}{ballester04}%
\bibitem{frey10a}%
  \BibitemOpen
  \bibfield{author}{%
  \bibinfo {author} {\bibfnamefont{M.}~\bibnamefont{Frey}}, \bibinfo {author}
  {\bibfnamefont{A.~L.}\ \bibnamefont{Miller}}, \bibinfo {author}
  {\bibfnamefont{L.~K.}\ \bibnamefont{Mentch}},\ and\ \bibinfo {author}
  {\bibfnamefont{J.}~\bibnamefont{Graham}},\ }%
  \bibfield{journal}{%
  \bibinfo {journal} {Quantum Information Processing}\ }%
  \textbf{\bibinfo {volume} {9}} (\bibinfo {month} {Mar}\ \bibinfo {year}
  {2010})%
  \bibAnnoteFile{NoStop}{frey10a}%
\bibitem{frey10b}%
  \BibitemOpen
  \bibfield{author}{%
  \bibinfo {author} {\bibfnamefont{M.}~\bibnamefont{Frey}}, \bibinfo {author}
  {\bibfnamefont{L.}~\bibnamefont{Coffey}}, \bibinfo {author}
  {\bibfnamefont{L.~K.}\ \bibnamefont{Mentch}}, \bibinfo {author}
  {\bibfnamefont{A.~L.}\ \bibnamefont{Miller}},\ and\ \bibinfo {author}
  {\bibfnamefont{S.}~\bibnamefont{Rubin}},\ }%
  \bibfield{journal}{%
  \bibinfo {journal} {Int. J. Quantum Information}}%
   (\bibinfo {year} {2010})%
  \bibAnnoteFile{NoStop}{frey10b}%
\bibitem{nielsen00}%
  \BibitemOpen
  \bibfield{author}{%
  \bibinfo {author} {\bibfnamefont{M.~A.}\ \bibnamefont{Nielsen}}\ and\
  \bibinfo {author} {\bibfnamefont{I.~L.}\ \bibnamefont{Chuang}},\ }%
  \emph{\bibinfo {title} {Quantum Computation and Quantum Information}}\
  (\bibinfo {publisher} {Cambridge University Press},\ \bibinfo {address}
  {Cambridge},\ \bibinfo {year} {2000})%
  \bibAnnoteFile{NoStop}{nielsen00}%
\bibitem{shapiro91}%
  \BibitemOpen
  \bibfield{author}{%
  \bibinfo {author} {\bibfnamefont{J.~H.}\ \bibnamefont{Shapiro}}\ and\
  \bibinfo {author} {\bibfnamefont{S.~R.}\ \bibnamefont{Shepard}},\ }%
  \bibfield{journal}{%
  \Doi{10.1103/PhysRevA.43.3795}{\bibinfo {journal} {Phys. Rev. A}}\ }%
  \textbf{\bibinfo {volume} {43}},\ \bibinfo {pages} {3795} (\bibinfo {month}
  {Apr}\ \bibinfo {year} {1991})%
  \bibAnnoteFile{NoStop}{shapiro91}%
\bibitem{king03}%
  \BibitemOpen
  \bibfield{author}{%
  \bibinfo {author} {\bibfnamefont{C.}~\bibnamefont{King}},\ }%
  \bibfield{journal}{%
  \Doi{10.1109/TIT.2002.806153}{\bibinfo {journal} {IEEE Trans. Inf. Theory}}\
  }%
  \textbf{\bibinfo {volume} {49}},\ \bibinfo {pages} {221} (\bibinfo {month}
  {Jan}\ \bibinfo {year} {2003})%
  \bibAnnoteFile{NoStop}{king03}%
\bibitem{ballester05}%
  \BibitemOpen
  \bibfield{author}{%
  \bibinfo {author} {\bibfnamefont{A.}~\bibnamefont{Dragan}}\ and\ \bibinfo
  {author} {\bibfnamefont{K.}~\bibnamefont{W\'odkiewicz}},\ }%
  \bibfield{journal}{%
  \Doi{10.1103/PhysRevA.71.012322}{\bibinfo {journal} {Phys. Rev. A}}\ }%
  \textbf{\bibinfo {volume} {71}},\ \bibinfo {pages} {012322} (\bibinfo {month}
  {Jan}\ \bibinfo {year} {2005})%
  \bibAnnoteFile{NoStop}{ballester05}%
\bibitem{slimen07}%
  \BibitemOpen
  \bibfield{author}{%
  \bibinfo {author} {\bibfnamefont{I.~B.}\ \bibnamefont{Slimen}}, \bibinfo
  {author} {\bibfnamefont{O.}~\bibnamefont{Trabelsi}}, \bibinfo {author}
  {\bibfnamefont{H.}~\bibnamefont{Rezig}}, \bibinfo {author}
  {\bibfnamefont{R.}~\bibnamefont{Bouall\`egue}},\ and\ \bibinfo {author}
  {\bibfnamefont{A.}~\bibnamefont{Boual\`egue}},\ }%
  \bibfield{journal}{%
  \bibinfo {journal} {J. Comp. Sci.}\ }%
  \textbf{\bibinfo {volume} {3}},\ \bibinfo {pages} {424} (\bibinfo {year}
  {2007})%
  \bibAnnoteFile{NoStop}{slimen07}%
\bibitem{boixo08}%
  \BibitemOpen
  \bibfield{author}{%
  \bibinfo {author} {\bibfnamefont{S.}~\bibnamefont{Boixo}}\ and\ \bibinfo
  {author} {\bibfnamefont{A.}~\bibnamefont{Monras}},\ }%
  \bibfield{journal}{%
  \Doi{10.1103/PhysRevLett.100.100503}{\bibinfo {journal} {Phys. Rev. Lett.}}\
  }%
  \textbf{\bibinfo {volume} {100}},\ \bibinfo {pages} {100503} (\bibinfo
  {month} {Mar}\ \bibinfo {year} {2008})%
  \bibAnnoteFile{NoStop}{boixo08}%
\bibitem{daems07}%
  \BibitemOpen
  \bibfield{author}{%
  \bibinfo {author} {\bibfnamefont{D.}~\bibnamefont{Daems}},\ }%
  \bibfield{journal}{%
  \Doi{10.1103/PhysRevA.76.012310}{\bibinfo {journal} {Phys. Rev. A}}\ }%
  \textbf{\bibinfo {volume} {76}},\ \bibinfo {pages} {012310} (\bibinfo {month}
  {Jul}\ \bibinfo {year} {2007})%
  \bibAnnoteFile{NoStop}{daems07}%
\bibitem{piani09}%
  \BibitemOpen
  \bibfield{author}{%
  \bibinfo {author} {\bibfnamefont{M.}~\bibnamefont{Piani}}\ and\ \bibinfo
  {author} {\bibfnamefont{J.}~\bibnamefont{Watrous}},\ }%
  \bibfield{journal}{%
  \Doi{10.1103/PhysRevLett.102.250501}{\bibinfo {journal} {Phys. Rev. Lett.}}\
  }%
  \textbf{\bibinfo {volume} {102}},\ \bibinfo {pages} {250501} (\bibinfo
  {month} {Jun}\ \bibinfo {year} {2009})%
  \bibAnnoteFile{NoStop}{piani09}%
\bibitem{helstrom67}%
  \BibitemOpen
  \bibfield{author}{%
  \bibinfo {author} {\bibfnamefont{C.~W.}\ \bibnamefont{Helstrom}},\ }%
  \bibfield{journal}{%
  \Doi{doi:10.1016/0375-9601(67)90366-0}{\bibinfo {journal} {Phys. Lett. A}}\
  }%
  \textbf{\bibinfo {volume} {25}},\ \bibinfo {pages} {101} (\bibinfo {month}
  {Jul}\ \bibinfo {year} {1967})%
  \bibAnnoteFile{NoStop}{helstrom67}%
\bibitem{braunstein94}%
  \BibitemOpen
  \bibfield{author}{%
  \bibinfo {author} {\bibfnamefont{S.~L.}\ \bibnamefont{Braunstein}}\ and\
  \bibinfo {author} {\bibfnamefont{C.~M.}\ \bibnamefont{Caves}},\ }%
  \bibfield{journal}{%
  \Doi{10.1103/PhysRevLett.72.3439}{\bibinfo {journal} {Phys. Rev. Lett.}}\ }%
  \textbf{\bibinfo {volume} {72}},\ \bibinfo {pages} {3439} (\bibinfo {month}
  {May}\ \bibinfo {year} {1994})%
  \bibAnnoteFile{NoStop}{braunstein94}%
\bibitem{paris09}%
  \BibitemOpen
  \bibfield{author}{%
  \bibinfo {author} {\bibfnamefont{M.}~\bibnamefont{Paris}},\ }%
  \bibfield{journal}{%
  \Doi{10.1142/S0219749909004839}{\bibinfo {journal} {Int. J. Quantum
  Information}}\ }%
  \textbf{\bibinfo {volume} {9}},\ \bibinfo {pages} {125} (\bibinfo {year}
  {2009})%
  \bibAnnoteFile{NoStop}{paris09}%
\bibitem{chen07}%
  \BibitemOpen
  \bibfield{author}{%
  \bibinfo {author} {\bibfnamefont{C.}~\bibnamefont{Ping}}\ and\ \bibinfo
  {author} {\bibfnamefont{L.}~\bibnamefont{Shunlong}},\ }%
  \bibfield{journal}{%
  \Doi{10.1007/s11464-007-0023-4}{\bibinfo {journal} {Front. Math. China}}\ }%
  \textbf{\bibinfo {volume} {2}},\ \bibinfo {pages} {359} (\bibinfo {year}
  {2007})%
  \bibAnnoteFile{NoStop}{chen07}%
\bibitem{frey09}%
  \BibitemOpen
  \bibfield{author}{%
  \bibinfo {author} {\bibfnamefont{M.}~\bibnamefont{Frey}}\ and\ \bibinfo
  {author} {\bibfnamefont{D.}~\bibnamefont{Collins}},\ }%
  \bibfield{journal}{%
  \bibinfo {journal} {Quantum Information and Computation VII, Proc. SPIE}\ }%
  \textbf{\bibinfo {volume} {7342}},\ \bibinfo {eid} {73420N} (\bibinfo {year}
  {2009})%
  \bibAnnoteFile{NoStop}{frey09}%
\bibitem{fujiwara99}%
  \BibitemOpen
  \bibfield{author}{%
  \bibinfo {author} {\bibfnamefont{A.}~\bibnamefont{Fujiwara}}\ and\ \bibinfo
  {author} {\bibfnamefont{P.}~\bibnamefont{Algoet}},\ }%
  \bibfield{journal}{%
  \Doi{10.1103/PhysRevA.59.3290}{\bibinfo {journal} {Phys. Rev. A}}\ }%
  \textbf{\bibinfo {volume} {59}},\ \bibinfo {pages} {3290} (\bibinfo {month}
  {May}\ \bibinfo {year} {1999})%
  \bibAnnoteFile{NoStop}{fujiwara99}%
\bibitem{horn85}%
  \BibitemOpen
  \bibfield{author}{%
  \bibinfo {author} {\bibfnamefont{R.~A.}\ \bibnamefont{Horn}}\ and\ \bibinfo
  {author} {\bibfnamefont{C.~R.}\ \bibnamefont{Johnson}},\ }%
  \emph{\bibinfo {title} {Matrix Analysis}}\ (\bibinfo {publisher} {Cambridge
  University Press},\ \bibinfo {address} {Cambridge},\ \bibinfo {year} {1985})%
  \bibAnnoteFile{NoStop}{horn85}%
\bibitem{giovannetti06}%
  \BibitemOpen
  \bibfield{author}{%
  \bibinfo {author} {\bibfnamefont{V.}~\bibnamefont{Giovannetti}}, \bibinfo
  {author} {\bibfnamefont{S.}~\bibnamefont{Lloyd}},\ and\ \bibinfo {author}
  {\bibfnamefont{L.}~\bibnamefont{Maccone}},\ }%
  \bibfield{journal}{%
  \Doi{10.1103/PhysRevLett.96.010401}{\bibinfo {journal} {Phys. Rev. Lett.}}\
  }%
  \textbf{\bibinfo {volume} {96}},\ \bibinfo {pages} {010401} (\bibinfo {month}
  {Jan}\ \bibinfo {year} {2006})%
  \bibAnnoteFile{NoStop}{giovannetti06}%
\end{thebibliography}

%Merlin.mbs v4.21 2009-07-09.
%

\end{document}